\documentclass[11pt,a4paper]{article}
\usepackage{jheppub}
\usepackage{graphicx,color}
\usepackage{amssymb}
\usepackage{multirow}

\newcommand{\beq}     {\begin{equation}}
\newcommand{\eeq}     {\end{equation}}
\newcommand{\bea}     {\begin{eqnarray}}
\newcommand{\eea}     {\end{eqnarray}}
\newcommand{\bit}     {\begin{itemize}}
\newcommand{\eit}     {\end{itemize}}
\newcommand{\ben}     {\begin{enumerate}}
\newcommand{\een}     {\end{enumerate}}
\newcommand{\bad}{\begin{array}{ccc}}
\newcommand{\ea}{\end{array}}

\newcommand{\bbt}{\bibitem}
\newcommand{\lsim}{\mathrel{\mathop{\kern 0pt \rlap
  {\raise.2ex\hbox{$<$}}}
  \lower.9ex\hbox{\kern-.190em $\sim$}}}
\newcommand{\gsim}{\mathrel{\mathop{\kern 0pt \rlap
  {\raise.2ex\hbox{$>$}}}
  \lower.9ex\hbox{\kern-.190em $\sim$}}}
  
\newcommand{\no}     {\nonumber}

\newcommand{\tb}      {{\tan\beta}}

\newcommand{\gm}      {\gamma}
\newcommand{\Gm}      {\Gamma}
\newcommand{\sg}      {\sigma}

\newcommand{\bb}      {{b \bar{b}}}

\newcommand{\rr}      {{\gamma\gamma}}
\newcommand{\ttau}      {{\tau\tau}}
\newcommand{\hsm}      {{h_{\rm SM}}}

\newcommand{\chtot}{{C^{ h}_{\rm tot}}}

\newcommand{\fig}{\begin{figure}}
\newcommand{\ef}{\end{figure}}

\newcommand{\ggf}{{gg {\rm F}}}
\newcommand{\vbf}{ {\rm VBF}}
\newcommand{\ggtt}{{gg {\rm F}+t\bar{t}h}}
\newcommand{\vvh}{{ {\rm VBF}+Vh}}
\newcommand{\fbi}{{\,{\rm fb}^{-1}}}

\newcommand{\gev}{{\;{\rm GeV}}}
\newcommand{\tev}{{\;{\rm TeV}}}

\def\lgeff{{\mathcal L}_{\rm eff}}

\newcommand{\wtR}{{\widetilde{R}\,}}

\newcommand{\yh}      { {\widehat{y}} }

\title{
Two Higgs doublet models for the LHC Higgs boson data at $\sqrt{s}=$ 7 and 8 TeV
}

\author[a]{Sanghyeon Chang,}
\author[b,c]{Sin Kyu Kang,}
\author[b]{Jong-Phil Lee,}
\author[d]{Kang Young Lee,}
\author[e]{Seong Chan Park,}
\author[f]{and Jeonghyeon Song}

\affiliation[a]{Faculty of Liberal Education, Seoul National University, Seoul, 151-742, Korea}
\affiliation[b]{School of Liberal Arts, Seoul-Tech, Seoul 139-743, Korea}
\affiliation[c]{PITT PACC, Department of Physics and Astronomy,
University of Pittsburgh, Pittsburgh, PA 15260, USA}
\affiliation[d]{Department of Physics Education,
Gyeongsang National University,
Jinju 660-701,
Korea}
\affiliation[e]{Department of Physics, Sungkyunkwan University,
Suwon 440-746, Korea}
\affiliation[f]{Division of Quantum Phases \& Devices, School of Physics,
Konkuk University, Seoul 143-701, Korea}

\abstract{
Updated LHC data on the new 126 GeV boson during the 7 and 8 TeV runnings 
strengthen the standard model Higgs boson interpretation further.
Through global $\chi^2$ fit analysis, we 
investigate whether the new particle could be  one of the scalar particles in two Higgs doublet models.
Four types of the model (I, II, X and Y) are comprehensively studied.
Considering the recent analysis on the spin-parity of the new boson,
we take two scenarios: it is either the lighter CP-even one  
or the heavier CP-even one. 
It is found that the current LHC Higgs data constrain the model quite strongly. 
Only narrow region along the decoupling line and a separate small island 
are allowed in Type II, X, and Y.
Type I is exceptional with much larger allowed space.
We also find that the current data are
compatible with the possibility that the light Higgs boson $h^0$ is hidden in the mass window of $90-100$ GeV.
}

\begin{document} 
\maketitle
\flushbottom

\section{Introduction}
\label{sec:introduction}
In July 2012, the ATLAS \cite{ATLAS-Higgs} and CMS \cite{CMS-Higgs} collaborations 
at the LHC
announced the discovery of a new boson
with mass around 126 GeV.
Both experiments had been looking for the Higgs boson in several decay channels, including
$\gamma\gamma, WW^{\ast}, ZZ^{\ast}, b\bar{b}$ and $\ttau$.
The signal rates in the $WW^{\ast}$ and $ZZ^{\ast}$ channels were in good agreement 
with the standard
model (SM) prediction, and those in the $b\bar{b}$ and $\ttau$ 
were also compatible 
with the SM.
However, there was an excess in the diphoton channel.
It was unclear whether the new boson is the long-sought SM Higgs boson or not.

Recently, the ATLAS and CMS have updated the Higgs search results  
using the full data recorded in 2011 and 2012 
with the integrated luminosity up to $5\fbi$ at 7 TeV \cite{ATLAS-7,CMS-7}
and $21\fbi$ at 8 TeV \cite{CMS-PAS-13-005,ATLAS-CONF-2013-034}.
The new data support the SM Higgs boson interpretation further,
even though each individual channel is still fluctuating.
For example, the excess in the diphoton channel decreased
in the updated CMS data, but retained in the ATLAS data:
\begin{eqnarray}
\mu_{\rr}=\left \{ \begin{array}{ll}
1.65^{+0.34}_{-0.30} & \hbox{ ATLAS;} \\[5pt]
0.78^{+0.28}_{-0.26} & \hbox{ CMS  (MVA mass-factorized);} \\[5pt]
1.11^{+0.32}_{-0.30} & \hbox{ CMS (Cut-based).}
\end{array} \right. \label{diphoton-13}
\end{eqnarray}

The current status is compactly encapsulated in a word ``a Higgs'', rather than ``the Higgs".
Even though the data seem to indicate very SM-like Higgs boson,
other scalar candidates in various new physics models are not excluded yet.
The quest for the identity of the new boson yields extensive studies in two directions.
One is global fit analysis in a model-independent way~\cite{baak,giardino,holdom,belanger,kingman}.
The other is to focus on a particular new physics model, and 
to place the constraints~\cite{np,brummer,arganda,hebecker,ibanez,ued}.

In this paper, we consider a minimal extension of the SM Higgs sector, 
a two Higgs doublet model (2HDM) with CP invariance.
Here are 5 scalar particles: 
CP-even light neutral Higgs $h^0$, CP-even heavy neutral Higgs $H^0$, 
CP-odd Higgs $A^0$, 
and charged Higgs $H^\pm$.
To suppress flavor changing neutral current (FCNC),
we assume a softly broken $Z_2$ symmetry.
According to the assignment of charges for quarks 
and leptons under the $Z_2$ symmetry, there are four types of 2HDM:
Type I, Type II, Type X, and Type Y
\cite{Glashow:Weinberg,2hdm:type1,2hdm:type2,2hdm:type3,Aoki,Ferreira}.
In the literature,
there are many studies about the implication of the LHC Higgs data 
on 2HDM \cite{Cheon,Gori,2013-2hdm,heavy,2hdcp,general2HDM}. 
Focused on Type II~\cite{type2}, or Type I and II~\cite{belanger},
the allowed parameter space has been obtained with 
electroweak precision constraints and flavor bounds.
The heavy Higgs search  is also studied in Refs.~\cite{Chen:2013rba,Craig:2013hca}.

In our previous work~\cite{CKLLPS}, 
we studied the implication of the early LHC Higgs data on 2HDM
in a comprehensive way.
In all of the four types of 2HDM,
we considered three possible scenarios consistent with the early LHC Higgs data.
With the latest LHC Higgs signals, 
we update the status of the 2HDM.
We pay attention to the spin-parity measurement of the new boson,
a very impressive step toward identifying it.
The angular distribution of four leptons 
in the $ZZ^*$ channel is compatible with the SM prediction 
$J^P=0^+$~\cite{ATLAS-spin,CMS-PAS-13-002}.
Other spin states like $J^P=0^-,~1^+,~1^-,~2^+$ are excluded at confidence levels (C.L.) above 97.8\%.
Considering this result,
we take two options:
the observed particle is either $h^0$ (Scenario-1) or $H^0$ (Scenario-2).

Our main questions are so as to how much parameter space of 2HDM
still survives especially outside the decoupling region,
whether 2HDM can explain the current data better than the SM, and
whether there is any chance to miss the light Higgs boson $h^0$ with $H^0$ being the observed one.
Intriguing is that the current LHC Higgs data with sizable uncertainties 
especially in fermonic decay modes constrain the model quite strongly.
Except for Type I model,
only a thin stripe survives outside the decoupling region.
The observation on \emph{multiple} Higgs decay channels 
is powerful in constraining new physics models.
Another unexpected result is
that the current LHC Higgs data start to 
predict the approximate characteristics of the hidden light Higgs boson $h^0$
in the Scenario-2.
Considering the null results in the LEP Higgs search,
the hidden $h^0$ is very likely in the mass range of $90-100\gev$.
As fitting the Higgs data to $H^0$,
the $h^0$ becomes very elusive at the LHC.
These are our main new results.

The paper is organized as follows.
We briefly review the 2HDM in Sec.~\ref{sec:review}.
Section \ref{sec:data} summarizes the latest LHC data on the Higgs signals.
In Sec.~\ref{sec:results}, the results of global $\chi^2$ fit analysis are given
for four types of 2HDM in two different scenarios.
Finally in Sec.~\ref{sec:conclusions} we conclude.
\section{Brief review of 2HDM}
\label{sec:review}
As one of the minimal extensions of the SM Higgs sector,
2HDM has two complex doublets of the Higgs fields:
\bea
\Phi_u = \left( \begin{array}{c}
             H_u^+ \\[5pt]
             \dfrac{v_u+H_u^0+iA_u^0}{\sqrt{2}} \end{array} \right )\;, \qquad
 \Phi_d = \left( \begin{array}{c}
             H_d^+ \\[5pt]
            \dfrac{v_d+H_d^0+iA_d^0}{\sqrt{2}}   \end{array} \right )\;.
\eea
Here $v_u$ and $v_d$ are non-zero vacuum expectation values (VEV),
which define the SM VEV via $v=\sqrt{v_u^2+v_d^2}$. 
The ratio of $v_u$ to $v_d$ is parametrized by an angle $\beta$ through 
$\tan\beta=v_u/v_d$.
Without loss of generality, we assume $\tb>0$.

Assuming CP invariance, there are five physical scalars,
the light CP-even scalar $h^0$,
the heavy CP-even scalar $H^0$, the CP-odd scalar $A^0$,
and two charged Higgs bosons $H^\pm$.
Physical states of neutral CP-even Higgs bosons are
\bea
h^0 = -H_d^0 \sin \alpha + H_u^0 \cos\alpha,  \quad
H^0 = H_d^0 \cos \alpha +H_u^0 \sin \alpha, 
\eea
where $\alpha$ is a mixing angle in the range of $[-\pi/2,\pi/2]$.

Yukawa interactions of $h^0$ and $H^0$ are parameterized by
\bea
 {\cal L}_{\rm Yuk} =
 - \sum_{f=u,d,\ell} \frac{m_f}{v}
 \left(
 \yh_f^h\bar{f}f h^0
 +
 \yh_f^H\bar{f}f H^0
 \right) .
\eea
In order to suppress FCNC,
we impose a discrete $Z_2$ symmetry in the Yukawa sector
so that one fermion couples 
with only one Higgs doublet.
There are four types of 2HDM with this discrete symmetry,
Type I, Type II, Type X, and
Type Y~\cite{2hdm:type1}.
The effective couplings of $\yh_f^{h,H}$ are referred to Ref.~\cite{CKLLPS}.
In the Higgs potential, however,
we allow a softly-broken $Z_2$-symmetric term, $-m_{12}^2 (\Phi_u^\dagger \Phi_d + H.c.)$.
The $m_{12}$ term plays an important role in naturally enhancing the charged Higgs boson mass.

Flavor physics significantly constrains the 2HDM parameters, especially $\tb$ 
and the charged Higgs boson mass.
Crucial observables are $b \to s \gamma$ and $\Delta M_{B_d}$,
which prohibit small $\tb$~\cite{flavor}.
The charged Higgs mass is required to be heavier than about $320\gev$ for Type II and Type Y.
For Type I and Type X, lighter $M_{H^\pm}$ is allowed.
Another observation with potential trouble to 2HDM is an excess of $B \to D \tau \nu$ events 
reported by the BaBar collaboration~\cite{BDtaunu:BaBar},
which contradicts the SM predictions
of lepton flavor universality.
The results cannot be accommodated in all four types of  2HDM with minimal flavor violation~\cite{BtoDtaunu}.
In the circumstance of no confirmation by the Bell experiment, we do not consider the effects here.
Finally we note that the constraint from $R_{b}$ in the electroweak precision data 
is weaker than those from flavor physics~\cite{Rbb}.

Considering the current LHC Higgs data and other constraints,
we study the following two scenarios:
\begin{description}
\item[Scenario-1:] The new boson $h$ is  $h^0$.
\item[Scenario-2:] It is $H^0$
while $h^0$ has been missed.
\end{description}
These do not include more exotic cases of two degenerate neutral Higgs bosons: 
a degenerate pair of $h^0$-$H^0$,
$h^0$-$A^0$, or $H^0$-$A^0$ may explain the LHC Higgs data~\cite{degenerate}.
Here
we focus on the normal setup.

There are eight free parameters in the general Higgs potential with CP invariance
and a softly broken $Z_2$ symmetry: the SM VEV $v$,
$m_{h^0}$, $M_{H^0}$, $M_{A^0}$, $M_{H^\pm}$, $m_{12}$,
$\alpha$, and $\tb$.
We assume heavy $m_{12}$, $M_{A^0}\simeq M_{H^\pm}$ with masses above 400 GeV.
The mass degeneracy between $A^0$ and $H^\pm$ is assumed for the suppression of new contributions to
the electroweak precision data \cite{Gunion:1989we,pdg}.
The other two masses, $m_{h^0}$ and $M_{H^0}$,
are determined according to the scenario type.
In the Scenario-1,
we put $m_{h^0}=126\gev$ while $M_{H^0} \geq 400\gev$.
In the Scenario-2, $m_{h^0} < M_{H^0} =126\gev$.
Remaining two free parameters are $\alpha$ and $\tb$.
From the flavor physics constraints,
we additionally constrain $\tb> 1.5~(\tb> 1)$ for Type I and Type X (Type II and Type Y)~\cite{CKLLPS}.
The upper bound on $\tb$ is set to be 50 for the
perturbativity \cite{Kanemura}.

The effective Lagrangian is \cite{Carmi:2012yp,Carmi:2012in}
\bea
\label{eq:Lg}
\lgeff &= &
c_V {2 m_W^2 \over v} h  \,  W_\mu^+ W_\mu^- + c_V  {m_Z^2 \over v} h  \, Z_\mu Z_\mu
\\ \no
&& - c_{b} {m_b \over v } h  \,  \bar b b  - c_{\tau} {m_\tau \over v } h \, \bar \tau \tau
- c_{c} {m_c \over v } h  \, \bar  c c
- c_{t} {m_t \over v } h  \, \bar  t t
\\ \nonumber &&
+ c_{g} {\alpha_s \over 12 \pi v} h  \, G_{\mu \nu}^a G^{a \mu \nu}
+ {c}_{\gamma } { \alpha \over \pi v} h  \, A_{\mu \nu} A^{\mu \nu}\,,
\eea
where $h=h^0$ in Scenario-1 and  $h=H^0$ in Scenario-2.
The SM values are 
$c_{V,\rm SM} =  c_{f,\rm SM} =1$, $c_{g,\rm SM}\simeq 1$
and $ {c}_{\gamma,\rm SM} \simeq -0.81$.
Without additional fermions or charged vector bosons,
$c_g$ and ${c}_\gm$ are determined by $c_{t,b,c,\tau,V}$.
The detailed expressions are in Ref.~\cite{CKLLPS}.

\section{Data on the LHC Higgs search and effective couplings for signals}
\label{sec:data}

As the Higgs data increase,
both ATLAS and CMS collaborations sort the results 
into two categories of production.
One is $\ggtt$, the combined results of the gluon fusion and the $t\bar{t}h$ production.
The other is $\vvh$ from the vector boson fusion ($\vbf$) and the associated production 
with $W$ or $Z$ gauge boson.
This classification is very efficient to understand the underlying physics since $\ggtt$ production is 
determined mainly by $t$-$\bar{t}$-$h$ vertex 
and $\vvh$ production by $V$-$V$-$h$ vertex.

A useful parameter is the ratio of the observed event rate of a specific channel to the SM expectation,
$R^{\tt production}_{\tt decay}$,
which is to be identified
with the signal strength modifier $\hat\mu = \sigma/\sg_{\rm SM}$.
In terms of the effective couplings, they are
\bea
\label{eq:R}
R_{\rr}^{gg \rm F}
&=&
\left|
\frac{c_g {c}_\gm}{{c}_{\gm,\rm SM} \chtot }
\right|^2,
\qquad
R_{ii}^{gg \rm F}
=
\left|\frac{c_g c_i}{ \chtot }\right|^2,
\\ \no
R_{ii}^{\vbf} &=& R_{ii}^{Vh}=R_{ii}^{\vvh}
=\left|\frac{c_V c_i }{ \chtot }\right|^2,
\\ \no
R_{\rr}^{\vbf}&=&R_{\rr}^{Vh}=R_{\rr}^{\vvh}
=
\left|
\frac{{c}_\gm c_V}{{c}_{\gm,\rm SM} \chtot }
\right|^2,
\eea
where $\chtot = \sqrt{\Gm_{\rm tot}^{h}/\Gm_{\rm tot}^{\hsm} }$, and
$i=W,Z,\tau,b$.

 {\renewcommand{\arraystretch}{1.4} 
\begin{table}
\centering
\caption{\label{tab:R}Summary of the LHC Higgs signals at 7 and 8 TeV.}
\begin{tabular}{|c|l|l|}
\hline
Production &  \multicolumn{1}{c|}{ATLAS}  & \multicolumn{1}{c|}{CMS}\\
\hline
$\ggtt$ & ~$\wtR_\rr^\ggtt = 1.47^{+0.66}_{-0.52}$ \cite{Jakobs} 
	      & ~$\wtR_\rr^\ggtt = 0.52 \pm 0.5$ \cite{CMS-PAS-13-001} \\
            & ~$\wtR_{WW}^\ggf = 0.82\pm0.36$ \cite{ATLAS-CONF-2013-030} 
	      & ~$\wtR_{WW}^\ggf = 0.73^{+0.22}_{-0.20}$ \cite{CMS-PAS-13-005} \\
            & ~$\wtR_{ZZ}^\ggtt = 1.8^{+0.8}_{-0.5}$ \cite{ATLAS-CONF-2013-013} 
	      & ~$\wtR_{ZZ}^\ggtt = 0.9^{+0.5}_{-0.4}$ \cite{CMS-PAS-13-002} \\
            & ~${\wtR_{\ttau}^\ggf = 1.0^{+2.1}_{-1.4}}$ \cite{ATLAS-CONF-2013-108} 
	      & ~${\wtR_{\ttau}^\ggf = 0.93 \pm 0.42 }$ \cite{1401.5041} \\
\hline
$\vvh$   & ~$\wtR_\rr^\vvh = 1.73^{+1.27}_{-1.11}$ \cite{Jakobs}~
	      & ~$\wtR_\rr^\vvh = 1.48 ^{+1.5}_{-1.1}$ \cite{CMS-PAS-13-001} \\
             & ~$\wtR_{WW}^\vbf = 1.66\pm0.79$ \cite{ATLAS-CONF-2013-030} 
	      & ~$\wtR_{WW}^\vbf = -0.05^{+0.75}_{-0.56}$, 
                     $\wtR_{WW}^{Vh} = 0.51^{+1.26}_{-0.94}$  \cite{CMS-PAS-13-005} \\
             & ~$\wtR_{ZZ}^\vvh = 1.2^{+3.8}_{-1.4}$ \cite{ATLAS-CONF-2013-013} 
	      & ~$\wtR_{ZZ}^\vvh = 1.0^{+2.4}_{-2.3}$ \cite{CMS-PAS-13-002} \\
             & ~${\wtR_{\ttau}^\vvh = 1.5^{+1.1}_{-1.0}}$ \cite{ATLAS-CONF-2013-108} 
	      &~${\wtR_{\ttau}^\vbf = 0.94\pm0.41 }$,
                     ${\wtR_{\ttau}^{Vh} = -0.33\pm1.02}$ \cite{1401.5041} \\
             & ~${\wtR_{\bb}^{\vvh} = 0.20 \pm 0.64}$ \cite{ATLAS-CONF-2013-079} 
	      & ~${ \wtR_{\bb}^{\vvh} = 0.96\pm 0.47 }$ \cite{1307.5745}
              \\ \hline
              \end{tabular}
\end{table}
}

In Table \ref{tab:R}, 
we summarize the observed
20 signal strengths $\wtR$,
reported by the ATLAS
and CMS collaborations
at the LHC with $\sqrt{s}=7\tev$
and $8\tev$.
Each individual signal strength explicitly shows that
there still exists some deviation 
from the SM expectation.

\section{Results of global fits to 2013 Higgs data}
\label{sec:results}
We perform global $\chi^2$ fits of model parameters to the observed Higgs signal strength
$\wtR_i$.
The $\chi^2$ is defined by
\bea
\label{eq:chi2:def}
\chi^2 = \sum_{i=1}^{20} \frac{(R_i - \wtR_i)^2}{\sigma_i^2},
\eea
where $i$ runs over all of the Higgs search channels and $\sigma_i$ is the uncertainty
of each channel.
For $\sigma_i$ we use the $1\sg$ systematic errors.

Global $\chi^2$ fits to the 20 data in Table \ref{tab:R} with the SM Higgs boson
hypothesis yield
\bea
\label{eq:chi2:sm}
\left.
\chi^2_{\rm SM}
\right|_{\rm d.o.f.=20} = 12.40.
\eea
Compared to 2012 data~\cite{CKLLPS},
the SM $\chi^2$ value is reduced.
This is mainly because of the reduction of $\rr$ mode measured by the CMS collaboration.
\subsection{Scenario-1}
\label{subsec:scenario-1}
The Scenario-1 is a normal setup such that the observed new scalar is the lightest CP-even Higgs boson
in 2HDM.
The effective couplings are
\bea
\label{eq:h:c}
c_V &=& \sin (\beta-\alpha),
\quad
c_b =\yh_d^{h},\quad
c_\tau = \yh_\ell^{h},
\quad
c_t =c_c = \yh_u^{h}.
\eea
Note that there exists the so-called decoupling limit where the light Higgs boson $h^0$
behaves exactly like the SM Higgs boson~\cite{decoupling}:
\bea
\hbox{Decoupling limit in Scenario-1: }
\sin (\beta-\alpha)=1.
\eea
In this limit, $c_V = c_f =1$.
The remaining free parameter, say $\tb$, 
does not affect the Higgs signals.

{\renewcommand{\arraystretch}{1.2} 
\begin{table}
\centering
\caption{\label{tab:scenario:1}The best-fit points 
and the corresponding couplings in Scenario-1. Note that $\chi^2_{\rm SM}/{\rm d.o.f}
= 0.62$.}
\vspace{1ex}
\begin{tabular}{|c||c|c|c||c|c|c|c|}
\hline
~Type~ & ~$\chi^2_{\rm min}/{\rm d.o.f}$~ & ~$\tb$~ &  ~$\cos(\beta-\alpha)$~ & ~$c_V$~ 
& ~$c_b$~ & ~$c_\tau$~ & ~$c_t$~  \\ \hline
I-1 & $0.58$ & $49.83$ & $0.42$ & $ 0.92$ & $0.92$ & $0.92$ & $0.92$ \\
II-1 & $0.64$ & $1.00$ & $-0.047$ & $1.00$ & $1.05$ & $1.05$ & $0.95$\\
X-1 & $0.60$ & $4.71$ & $0.40$ & $0.92$ & $1.00$ & $-0.97$ & $1.00$ \\
Y-1 & $0.62$ & $4.94$ & $0.40$ & $0.92$ &$-1.06$ & $1.00$ & $1.00$ \\
\hline
              \end{tabular}
\end{table}
}

We perform global $\chi^2$ fits to the new LHC Higgs data,
and find the $\chi^2$ minimum point for each type.
In order to compare the SM results,
we present the $\chi^2_{\rm min}$ per degree of freedom (d.o.f.)
in 
Table \ref{tab:scenario:1}.
Note that 2HDM with two free parameters 
has 18 d.o.f. while the SM has 20.
As the $\chi^2_{\rm min}/{\rm d.o.f}$ values imply,
all of the best-fit points are as good as the SM in explaining the Higgs data.
Type I best-fit point has the smallest $\chi^2_{\rm min}/{\rm d.o.f}$, 
although not significantly improved from the SM.
Considering the presence of the decoupling limit in the 2HDM,
this compatibility is not surprising.
Interesting is that
the best-fit points in Type I, X, and Y
are located away from the decoupling limit,
as indicated by $\cos(\beta-\alpha) \simeq 0.4$.
Their effective couplings show some deviation from the SM values.
At the Type I best-fit point, all of the effective couplings
are smaller than the SM ones by about 8\%.
Type X best-fit point has only one sizable deviation in $c_V$.
For the Type Y, the effective couplings of $c_V$ and $c_b$ are about 10\% different.
On the while, Type II best-fit point is practically the same as the SM.

\begin{figure}[tbp]
\centering 
\includegraphics[width=.47\textwidth]{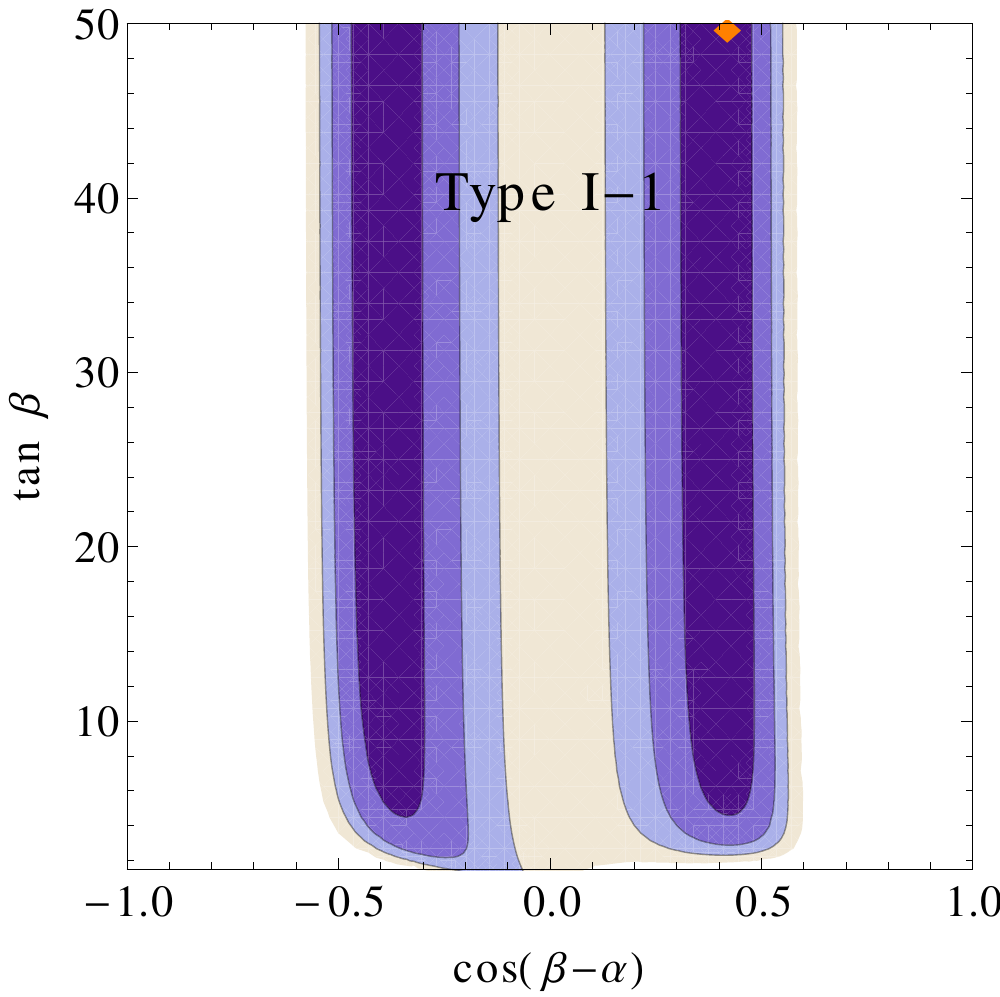}
~~
\includegraphics[width=.48\textwidth]{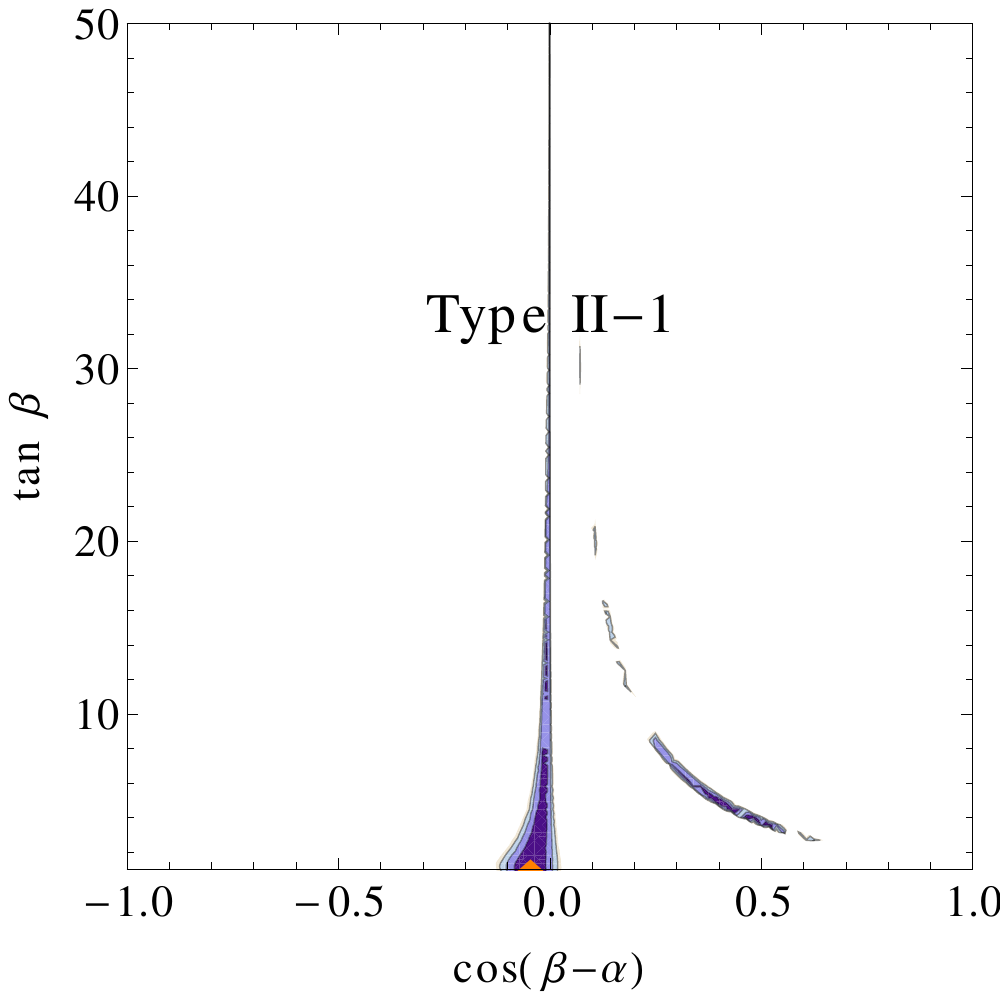}
\\
\includegraphics[width=.47\textwidth]{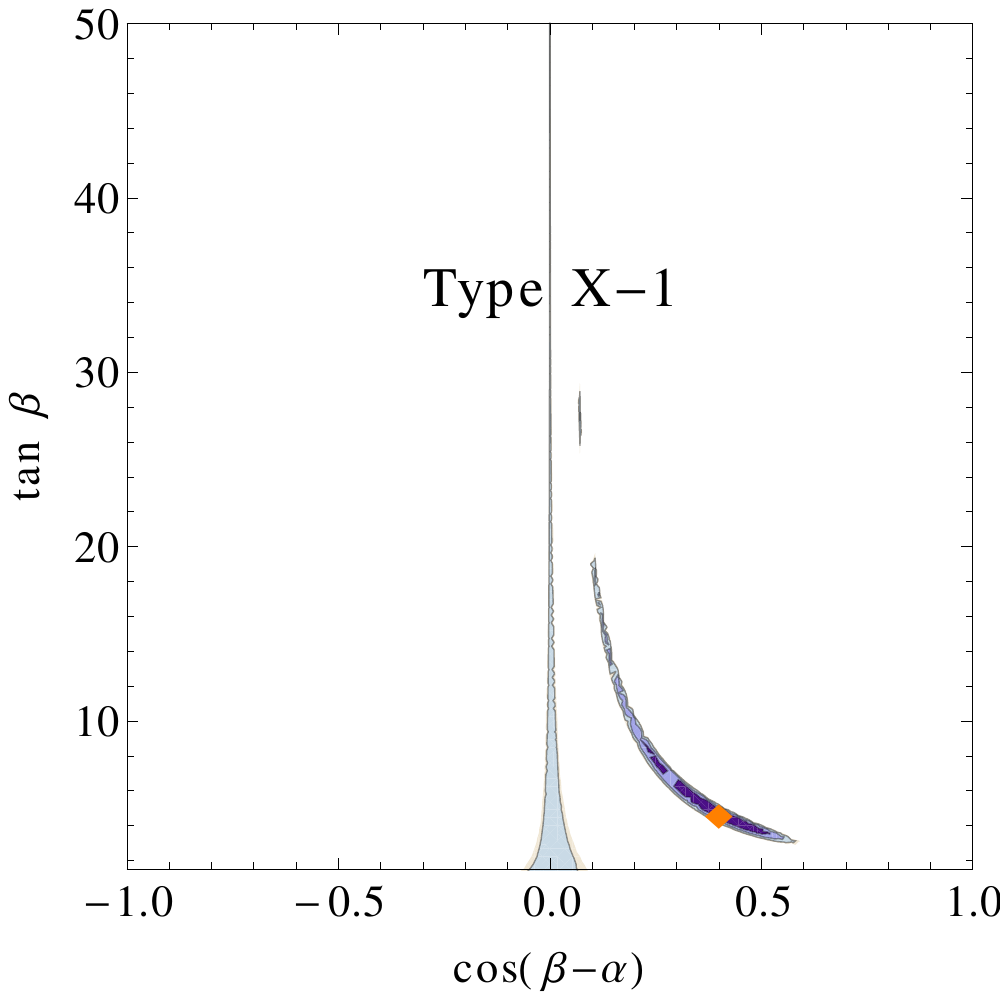}
~~
\includegraphics[width=.47\textwidth]{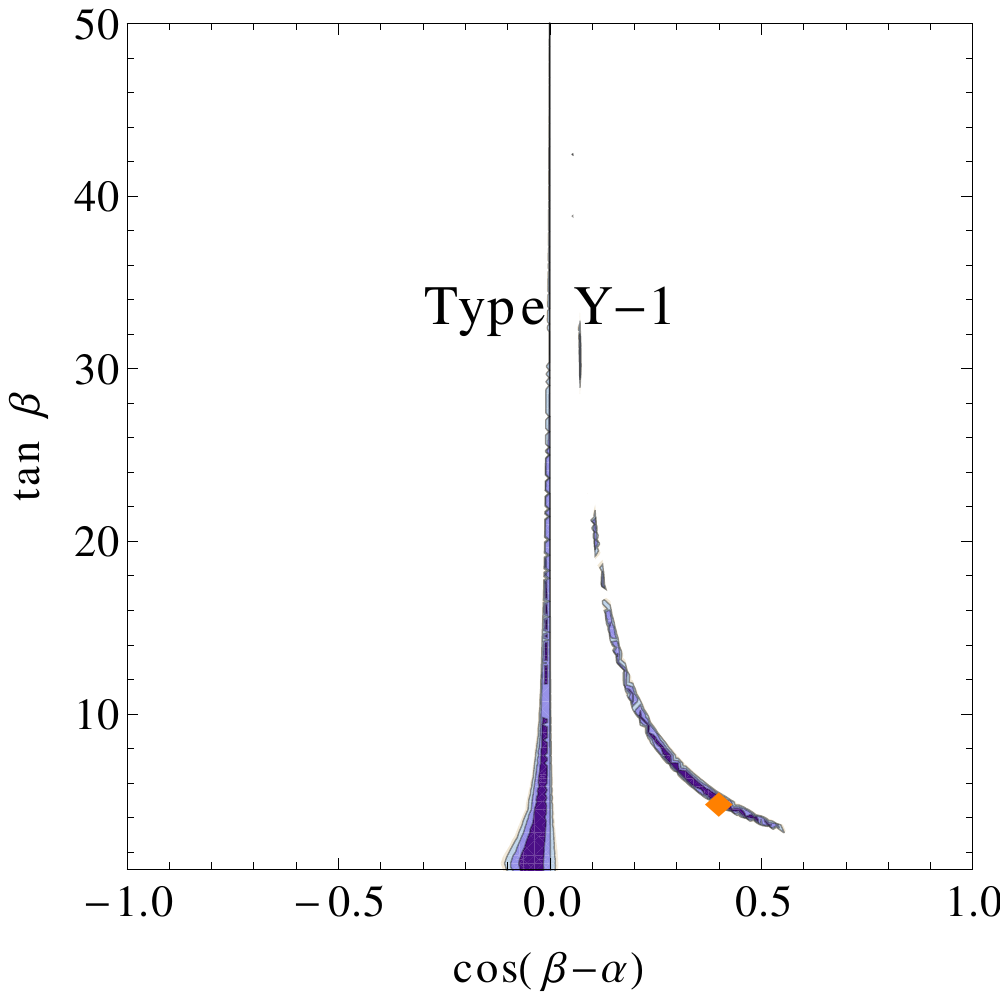}
\caption{\label{fig:scenario:1}Allowed
regions of the Scenario-1 at $1\sg$ in the parameter space of $(\cos(\beta-\alpha),\tb)$
for Type I, Type II, Type X, and Type Y models.
The darker the region is, the smaller the $\chi^2$ value is.
The decoupling limit is along the central line, $\cos(\beta-\alpha)=0$.
Orange diamonds denote the best-fit points for each type.
}
\end{figure}

Brief comments on negative Yukawa couplings~\cite{negative} are in order here.
At the best-fit points,
$c_\tau$ in Type X and $c_b$ in Type Y
become negative.
Both best-fit points
are located in the positive $\alpha$ region, away from the decoupling line.
Since $c_\tau$ in Type X and $c_b$ in Type Y
are $-\sin\alpha/\cos\beta$,
they become negative ($\beta$ is defined as a positive angle).
In order to see which observables in Table \ref{tab:R}
prefer these negative Yukawa couplings,
we perform the global $\chi^2$ fit only in the $\alpha<0$ (equivalently $c_f>0$)
region, find the best-fit point,
and compare each $\chi^2$ based on 20 observables
with that from the true best-fit point.
For positive Yukawa coupling, the $\chi^2_{\rm min}/{\rm d.o.f.}$ value is increased:
for Type X, the increase is 13.3\%;
for Type Y, it is 1.5\%.
The preference to negative $c_\tau$ in Type X is 
attributed to the CMS reduced rates of $\wtR_\rr^\ggtt$,
$\wtR_{WW}^\ggf$, and 
$\wtR_{WW}^\vbf$: see Table \ref{tab:R}.
With negative $c_\tau$,
the $\tau$ contribution to the diphoton rate has 
the same sign with the $W$ contribution,
which allows smaller $c_V$.
The reduced $\wtR_{WW}^\ggf$ and
$\wtR_{WW}^\vbf$ become more consistent.
In Type Y, 
the $b$ quark has one third charge of $\tau$,
of which the effect is smaller.

Another question is whether we can observe this negativeness in the Higgs data.
The observation requires the interference with other diagrams having positive couplings.
Among the Higgs decay channels, 
loop-induced ones like $\rr$, $gg$, and $Z\gm$ 
are able to probe this interference.
But this requires very high precision since the contribution of $c_\tau$ or $c_b$
are minor.
The dominant contributions to the $\rr$ mode, for example,
are from $W$ and top loop. Both effective couplings 
have the same sign in this case.
Future linear colliders 
like the ILC~\cite{ILC}, TLEP \cite{TLEP} and the muon collider Higgs factory \cite{MCHF}
are expected to perform this observation.

Although the best-fit point is the most probable in the given model,
the degree of its credibility should be answered statistically.
Particularly when the $\chi^2$ plot is shallow along a specific parameter,
we cannot insist on the best-fit point only.
This is the case for the decoupling limit:
once $\sin (\beta-\alpha)=1$,
the value of $\tb$ does not affect the Higgs phenomenology;
the $\chi^2$ plot against $\tb$ is generically shallow along the decoupling line.

In Fig.~\ref{fig:scenario:1}, we show the allowed region at $1\sg$
in the parameter space of $(\cos(\beta-\alpha), \tb)$
by the 2013 LHC Higgs data.
The darker the region is, the lower the $\chi^2$ value is.
The decoupling limit is along the central line, {\it i.e.,} $\cos(\beta-\alpha)=0$.
The most important conclusion is that
except for Type I the current LHC Higgs data constrain the 2HDM
quite strongly.
The allowed regions for Type II,  X and Y,
of which the shape and location look alike to each other,
are very limited.
Along the decoupling limit,
only a narrow band remains.
Away from it,
most of the parameter space is excluded at $1\sg$
except for an island group of the shape of a short ribbon.
Minor difference is in Type-X, where the island region is clearly favored.
Type II and Type Y prefer the decoupling region and the island region almost equally.
 
Type I is exceptional.
The allowed region at $1\sg$ is much more widespread than those of the other three types.
A large portion of the parameter space is still consistent with the current Higgs data.
In addition, the most preferred (darkest)
region near the $\chi^2$ minimum point is not along the decoupling limit.
It is of a long stripe shape with $\cos(\beta-\alpha)\simeq \pm 0.4$ and $\tb\gsim 5$.

\begin{figure}[tbp]
\centering 
\includegraphics[width=.70\textwidth]{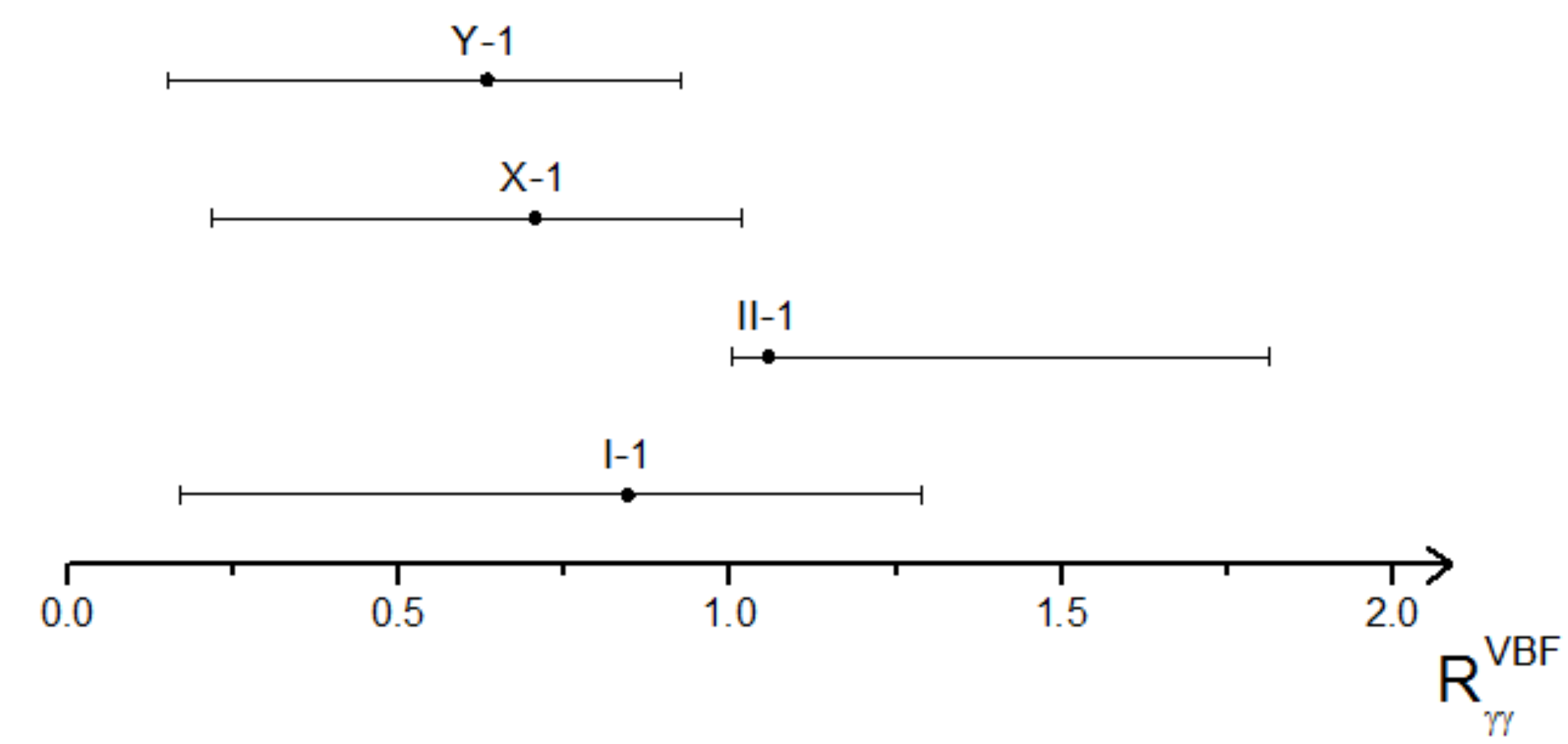}
\caption{\label{fig:scenario:1:R} 
Signal strength $R^\vbf_\rr$ for models of I-1, II-1, X-1, and Y-1
with $1\sg$.
The black blobs are the predictions of the best-fit point.
}
\end{figure}

The next question is whether we can distinguish each type
from the LHC Higgs data.
This may be answered by comparing the signal strengths 
at four best-fit points.
We find that
the signal strengths are different with variance up to 50\%.
The most efficient signal is $R^\vbf_\rr$,
which is about 1.1 for Type II, 0.7 for Type I, 0.4 for Type X and 0.3 for Type Y.
However, the best-fit point is under statistical uncertainty.
In Fig.~\ref{fig:scenario:1:R},
we show the $R^\vbf_\rr$ values with $1\sg$ uncertainty.
The best-fit point predictions are marked by dots,
which are quite different.
With $1\sg$ uncertainty, however,
all of the four types are overlapped.
We need much higher precision to probe the differences among different types of 2HDM.

\subsection{Scenario-2}

The Scenario-2 is rather exotic such that the light $h^0$ has not been observed yet
and the observed new boson is the heavy CP-even $H^0$.
The effective couplings are then
\bea
\label{eq:H:c}
c_V &=& \cos (\beta-\alpha),
\quad
c_b =\yh_d^{H},\quad
c_\tau = \yh_\ell^{H},
\quad
c_t =c_c = \yh_u^{H}.
\eea

In order to evade the LEP Higgs search~\cite{LEP,LEP2},
we demand that the event rate of flavor-independent jet decay of the light Higgs boson $h^0$
be smaller than the observed limit.
This rate $|\xi|^2$ is the most strongly constrained one.
In terms of the effective couplings, it is
\bea
|\xi|^2
=
|c_V|^2 \cdot
\frac{ {\cal B}(h^0 \to jj) }{{\cal B}(\hsm \to jj)} .
\eea
$|\xi|^2$ depends on the $h^0$ mass.
We examine whether there is an additional resonance peak
in the diphoton invariant mass distribution at the LHC.
In the early LHC data, the distribution started from $110\gev$.
In 2013 data, it is presented from $100\gev$.
Since there is no sign of a resonance in the low energy region, 
we take a conservative stance to assume  $m_{h^0}=90\gev$.
The LEP upper bound is then  $|\xi|^2 < 0.155$ \cite{LEP}.

{\renewcommand{\arraystretch}{1.2} 
\begin{table}
\centering
\caption{\label{tab:scenario:2}The best-fit points and the corresponding couplings in Scenario-2.}
\begin{tabular}{|c||c|c|c||c|c|c|c|}
\hline
~~~Type~~~ & ~~$\chi^2_{\rm min}/{\rm d.o.f}$~~ & ~~$\tb$~~ & ~~~$\sin(\beta-\alpha)$~~~ & ~~~$c_V^H$~~~  
& ~~~$c_b^H$~~~ & ~~~$c_\tau^H$~~~ & ~~~$c_t^H$~~~  \\ \hline
I-2 & $0.58$ & $50.0$ & $0.40$ & $ -0.92$ & $-0.93$ & $-0.93$ & $-0.93$ \\
II-2 & $0.59$ & $50.0$ & $3 \times 10^{-4}$ & $1.00$ & $1.01$ & $1.01$ & $1.00$ \\
X-2 & $0.60$ & $4.72$ & $0.40$ & $-0.92$ & $-1.00$ & $0.97$ & $-1.00$ \\
Y-2 & $0.59$ & $50.0$ & $3 \times 10^{-4}$ & $1.00$ & $1.01$ & $1.00$ & $1.00$ \\
\hline
              \end{tabular}
\end{table}
}

\begin{figure}[tbp]
\centering 
\includegraphics[width=.47\textwidth]{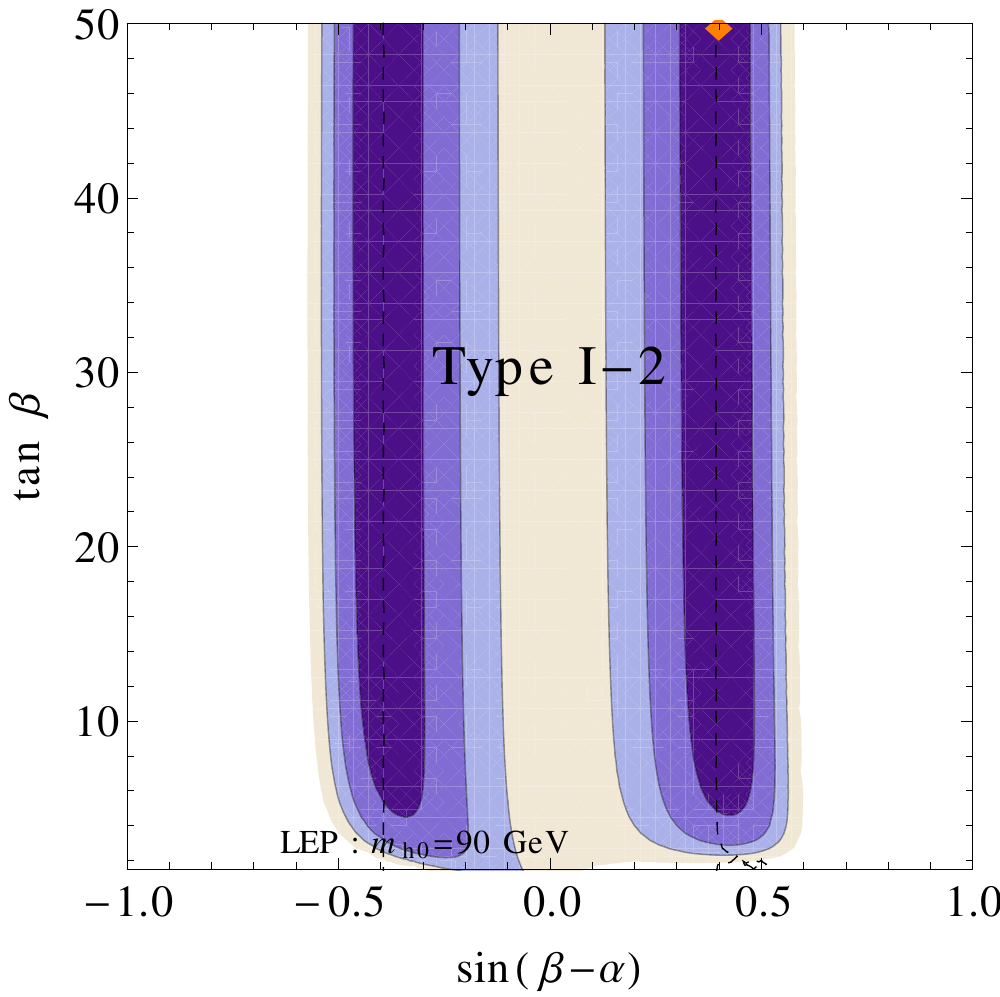}
~~
\includegraphics[width=.47\textwidth]{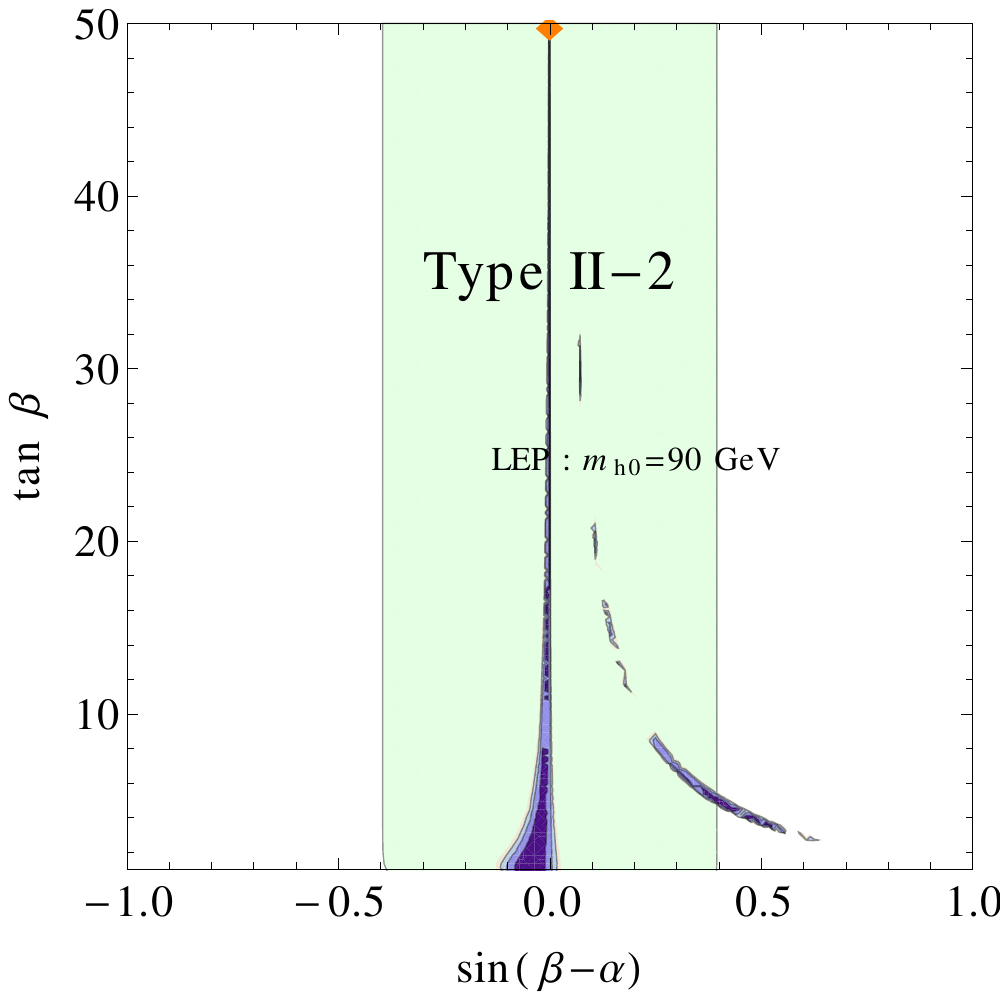}
\\
\includegraphics[width=.47\textwidth]{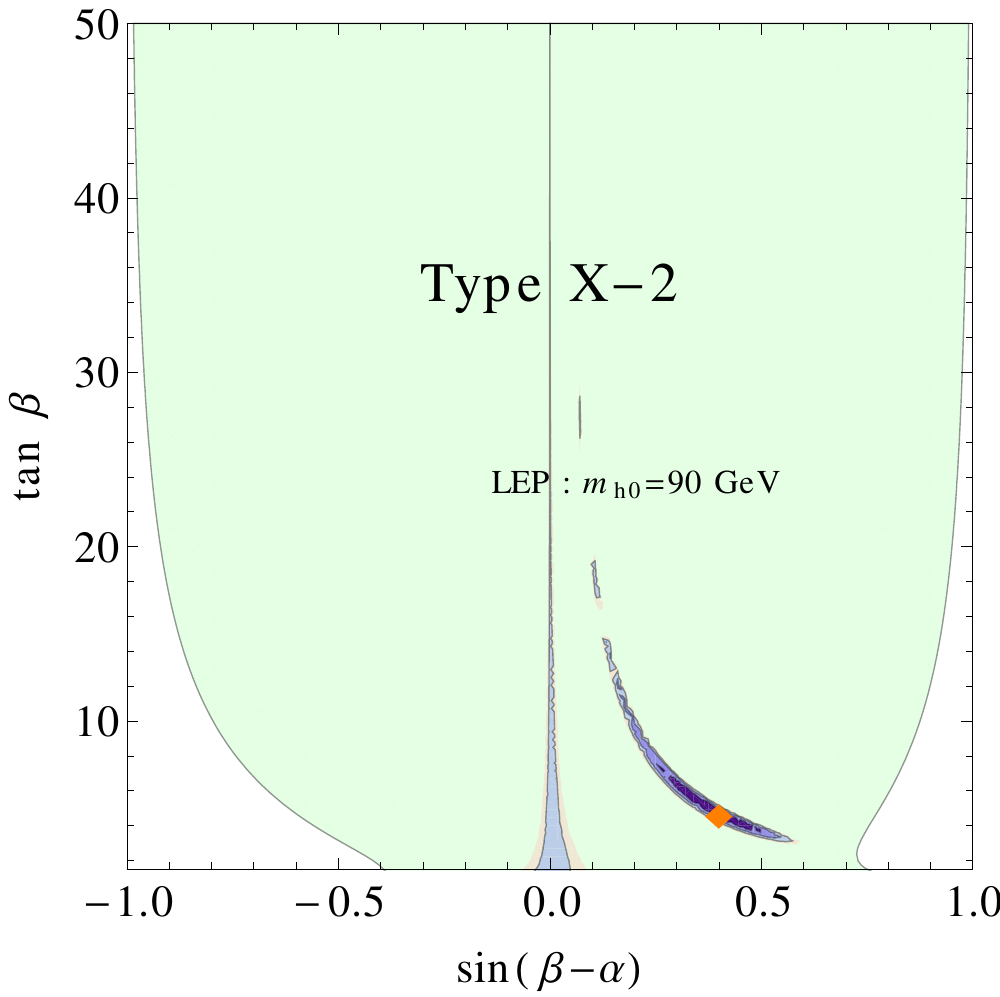}
~~
\includegraphics[width=.47\textwidth]{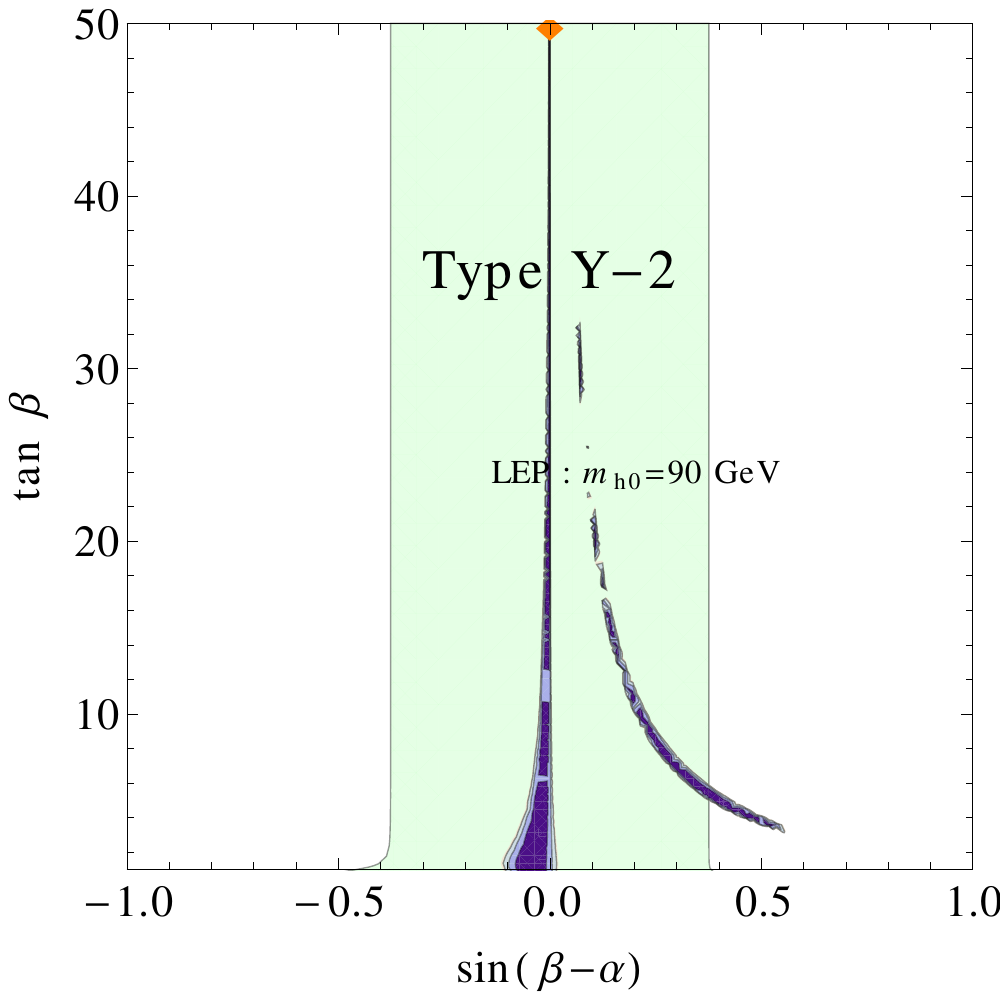}
\caption{\label{fig:scenario:2} Allowed regions at $1\sg$
by the current LHC Higgs data for the Scenario 2
where the observed $126\gev$ boson the the Heavy Higgs boson $H^0$.
The brighter region is allowed by the LEP bound for the case of $m_{h^0}=90\gev$.
Orange diamonds denote the best-fit points for each type.
}
\end{figure}

In the parameter space allowed by perturbativity, flavor physics,
and the LEP bound,
we perform global $\chi^2$ fits, and find the best-fit points.
The best-fit points including their effective couplings are summarized in Table \ref{tab:scenario:2}.
In Fig.~\ref{fig:scenario:2},
we present the $1\sg$ allowed parameter space of $(\sin(\beta-\alpha),\tb)$
in the Scenario-2.
The darker the allowed region is, the smaller the $\chi^2$ value is. 
The light green region is the LEP bound for the case of $m_{h^0}=90\gev$.
If $m_{h^0}$ increases, the LEP bound gets weaker.
The pattern of the allowed region for each type in the Scenario-2 
is very similar to that in the Scenario-1.
This is because of the relation $\left.\alpha\right|_{\rm Scenario-2} +\pi/2 
= \left.\alpha\right|_{\rm Scenario-1}$.
Unexpected is that the LEP bound on the light Higgs boson is rather weak.
The constraints from the LHC Higgs data are stronger.
For Type Y, however, the LEP bound excludes the LHC Higgs best-fit point
around $\sin(\beta-\alpha) \simeq 0.4$.

There are minor differences between Scenario-1 and Scenario-2.
In Type II and Type Y of the Scenario-2, 
the $\chi^2_{\rm min}$ values get a little bit improved than in those for the Scenario-1.
It is attributed to the additional decay mode  $H^0 \to h^0 h^{0*} \to \bb\bb$.
In most parameter space, its branching ratio is negligible.
Exceptions occur in the decoupling limit for Type II and Type Y,
where the $h^0$-$b$-$\bar{b}$ couplings become proportional to $\tb$~\cite{largeHhh}.
For the large value of $\tb$,
therefore,  $H^0 \to 4b $ mode becomes non-negligible,
of which the maximum branching ratio is about 10\% for Type II.
This additional decay mode increases the total decay width $\Gm^h_{\rm tot}$ in Eq.(\ref{eq:R}).
Since our model predicts smaller $R$ values compared to the observed $\wtR$,
$\chi^2$ defined in Eq.(\ref{eq:chi2:def}) decreases with increasing $\Gm^h_{\rm tot}$.

{\renewcommand{\arraystretch}{1.2} 
\begin{table}
\centering
\caption{\label{tab:scenario:2:h}The best-fit points and the corresponding couplings of the
light CP-even Higgs boson with mass $m_h =90\gev$ in Scenario-2.}
\begin{tabular}{|c||c|c|c|c|}
\hline
~Type~ & ~~I-2~~ & ~~II-2~~ & ~~X-2~~ &  ~~Y-2~~ \\ \hline \hline
$R_\rr^\ggf$ &  $0.15$  &  $4.5\times 10^{-3}$ & $4.0\times 10^{-3}$ & $9.0\times 10^{-4}$\\
$R_\rr^\vbf$ & $0.18$ &$1.9\times 10^{-11}$ & $1.6\times 10^{-2}$ & $3.7 \times 10^{-12}$
\\
\hline
              \end{tabular}
\end{table}
}

In order to confirm the elusiveness of the light CP-even Higgs boson,
 we present the signal strengths
$R_\rr^\ggf$ and $R_\rr^\vbf$ in Table \ref{tab:scenario:2:h}.
For all types of 2HDM,
the diphoton signals are negligible.
The couplings with the gauge boson, $c_V$,
are all much smaller than the SM one.
At the LHC, the observation of this resonance in the diphoton mode is very unlikely.

%
\section{Conclusions}
\label{sec:conclusions}
We have updated the status of CP-conserving 2HDM with a softly-broken $Z_2$ symmetry,
based on the latest LHC Higgs data.
Four types of models are comprehensively investigated.
Accepting the new spin-parity measurement of $J^P = 0^+$,
we consider two scenarios where the observed $126\gev$ particle is 
the light CP-even Higgs $h^0$ (Scenario-1) or the heavy CP-even $H^0$ (Scenario-2).
We have found that in both scenarios 
the current LHC Higgs data constrain 2HDM quite strongly.
The decoupling region, which should be allowed by the SM Higgs-like data,
is also very limited. 
Away from the decoupling limit, 
most parameter space is excluded except for a small island region.
One exception is Type I.
A large portion of the parameter space is allowed at $1\sg$.
And the best-fit point is apparently separated from the decoupling line.

An interesting possibility is the Scenario-2:
the observed new particle is the heavy CP-even Higgs $H^0$ of the 2HDM 
while the light CP-even Higgs $h^0$ is buried in the mass window of  $90-100\gev$.
Since the Higgs phenomenology in the Scenario-2 is the same as
that in the Scenario-1 if $\alpha \to \alpha+\pi/2$,
the presence of the similar allowed parameter space is expected.
Unexpected is that the LEP Higgs search bound is rather weak.
It is very likely that all of the four types of 2HDM models
may survive with larger LHC data in the future.

\acknowledgments
This work was supported in part by the National Research Foundation of
Korea (NRF) grant funded by the Korea government of the Ministry of Education, Science
and Technology (MEST) (No. 2011-0003287).
The work of JS is supported by NRF-2013R1A1A2061331.
K.Y.L. was supported by the Basic Science Research Program through the NRF funded by MEST (2010-0010916).
S.C.P. is supported by NRF-2013R1A1A2064120,
and Basic Science Research Program through the NRF of Korea
funded by the MEST
(2011-0010294) and (2011-0029758).

\end{document}